\documentclass[aps,prb,superscriptaddress,twocolumn,showpacs]{revtex4-1}

\usepackage{amsmath, amsthm, amssymb}
\usepackage{graphicx}
\usepackage{color}

\usepackage{array}
\newcolumntype{V}{>{\centering\arraybackslash} m{.7\linewidth} }

\usepackage{hyperref}
\hypersetup{
        colorlinks=true
}

\newcommand{\eqnref}[1]{(\ref{#1})}

\newcommand{\ceff}{c_{\text{eff}}}

\usepackage{trsym}
\usepackage{relsize}
\newcommand{\up}{\mathsmaller{\InversTransformVert}}
\newcommand{\dw}{\mathsmaller{\TransformVert}}

\long\def\beginpgfgraphicnamed#1#2\endpgfgraphicnamed{\includegraphics{#1}}

\begin{document}

\title{A supersymmetric multicritical point in a model of lattice fermions}

\author{Bela Bauer}
\affiliation{Station Q, Microsoft Research, Santa Barbara, CA 93106-6105, USA}

\author{Liza Huijse}
\affiliation{Department of Physics, Harvard University, Cambridge MA 02138}

\author{Erez Berg}
\affiliation{Department of Condensed Matter Physics, Weizmann Institute of Science, Rehovot 76100, Israel}

\author{Matthias Troyer}
\affiliation{Theoretische Physik, ETH Zurich, 8093 Zurich, Switzerland}

\author{Kareljan Schoutens}
\affiliation{Institute for Theoretical Physics, University of Amsterdam, Science Park 904, P.O.Box 94485, 1090 GL Amsterdam, The Netherlands}

\begin{abstract}

We study a model of spinless fermions with infinite nearest-neighbor repulsion on the square ladder which has microscopic supersymmetry. It has been conjectured that in the continuum the model is described by the superconformal minimal model with central charge $c=3/2$. Thus far it has not been possible to confirm this conjecture due to strong finite-size corrections in numerical data. We trace the origin of these corrections to the presence of unusual marginal operators that break Lorentz invariance, but preserve part of the supersymmetry. By relying mostly on entanglement entropy calculations with the density-matrix renormalization group, we are able to reduce finite-size effects significantly. This allows us to  unambiguously determine the continuum theory of the model. We also study  perturbations of the model and establish that the supersymmetric model is a multicritical point. Our work underlines the power of entanglement entropy as a probe of the phases of quantum many-body systems.

\end{abstract}



\maketitle

\section{Introduction}

The study of supersymmetry has its origins in the field of particle physics, where it allows the construction of quantum field theories containing both bosonic and fermionic excitations which are related by the supersymmetry. Such field theories are considered promising candidates for extensions of the standard model of particle physics and may solve some of the open problems encountered in the field today.

At the same time, supersymmetry has found applications in many other fields of physics.
In this paper, we study a supersymmetric model for spinless interacting fermions in one spatial dimension which was first introduced in Refs.~\onlinecite{fendley2003,fendley2003-1}. Our motivation for studying this model is twofold: On the one hand, the supersymmetry allows us to make rigorous statements despite the strongly interacting nature of the model. We will see that the model displays a multicritical point and that supersymmetry allows us to easily identify this point in parameter space. Generically, identifying and tuning to such a multicritical point is difficult. Imposing the exact supersymmetry allows us to numerically study the vicinity of the critical point and establish ties to the continuum theory.

On the other hand, this model allows us to benchmark and improve the numerical tools used to explore criticality in one dimension. We show that the critical theory for this model contains marginal operators which give rise to strong finite-size corrections. Such corrections often cause severe obstacles to numerically approaching critical systems. We will show how these obstacles can be overcome by relying on the entanglement entropy, which is less affected by finite-size corrections.

The model we study falls in a broader class of models for interacting fermions with supersymmetry. In these models the degrees of freedom are spinless fermions with local repulsive interactions. There are no microscopic bosonic degrees of freedom, as one might have expected for a supersymmetric theory. Instead supersymmetry relates fermionic and bosonic many-body states, i.e. states with an odd or even number of fermions, respectively. The Hamiltonian contains a nearest-neighbor hopping term and interactions for particles less than two sites apart. Imposing supersymmetry fixes the strength of the terms in the Hamiltonian, but -- as we will discuss below -- it was found that a rich array of phases can be found when the underlying lattice is varied.

We now review some recent results on this class of models to illustrate the power of incorporating supersymmetry in models for interacting fermions. In Refs.~\onlinecite{fendley2003,fendley2003-1}, the model was proposed and solved explicitly on a chain using the Bethe ansatz. Exploiting the additional tools that are available due to supersymmetry, further aspects of these models were understood: Using the Witten index and cohomology arguments, the number of ground states and some of their properties on several two-dimensional lattices were obtained in Refs.~\onlinecite{fendley2005,huijse2008}. This was used to demonstrate that the models exhibit surprising properties such as superfrustration,\cite{fendley2005} an exponential degeneracy of the ground state which is not lifted by quantum fluctuations. The case of the chain was analyzed in more detail in Ref.~\onlinecite{beccaria2005}, where the authors explore properties of the ground state for finite systems. The spectrum and its relation to superconformal field theory was explored in detail in Ref.~\onlinecite{huijse2011}. Recent advances include the study of perturbations of the model which preserve supersymmetry, namely staggered interactions.\cite{fendley2010,fendley2011,huijse2011-1,beccaria2012} Interesting extensions of the supersymmetric model revealing relations to various spin chains were also explored.\cite{fendley2003-1, hagendorf2012}

In this paper, we focus on this model on the square ladder. This model turns out to pose a particularly interesting numerical challenge. It has been shown to be critical using the behavior under boundary twists.~\cite{huijse2008} It has been conjectured that its continuum theory is the second $\mathcal{N}=2$ superconformal minimal model,\cite{fendley2003} but a reliable numerical confirmation of this conjecture has not been achieved so far due to strong finite-size effects in exact diagonalization calculations. In general, finite-size effects are a result of the presence of irrelevant and marginally irrelevant operators in the ultraviolet (UV) theory. Marginally irrelevant operators are particularly notorious as they lead to corrections that are suppressed only logarithmically in the system size. One might expect that fine-tuning the microscopic model to be supersymmetric would prevent such problems from arising. We will see, however, that this is not the case, and that the supersymmetry imposed on the lattice still allows for the presence of a marginal operator. We resolve this problem by realising that finite-size corrections are strongly suppressed in the entanglement properties. We study the entanglement properties of the system at the supersymmetric point and a number of perturbations away from this point, which allow us to establish the phase diagram and identify the supersymmetric point as the multicritical point in that phase diagram. We thus confirm numerically that the continuum limit of the supersymmetric model on the square ladder is described by the second superconformal minimal model.

The outline of this paper is as follows: In Section~\ref{sct:model}, we review the construction of the supersymmetric lattice model for hard-core fermions with a focus on the relevant properties of the model on the square ladder. In Section~\ref{sct:scft}, we review the conjectured continuum theory for this model and discuss the origin of the large finite-size corrections.

In Section~\ref{sct:methods}, we turn our attention to the numerical methods that will be used to identify the continuum theory and study the phase diagram. Our numerical results are presented in Section~\ref{sct:results}. First we present results for the supersymmetric point (Sct.~\ref{sct:resA}), then we discuss perturbations away from the critical point that allow us to explore the two-dimensional phase diagram surrounding the supersymmetric point (Sct.~\ref{sct:pd}). Finally, results for a supersymmetry-preserving perturbation of the supersymmetric model are presented in Sct.~\ref{sct:zstag}.

In Appendix~\ref{app:sollimits}, we review previous work on exactly soluble limits of the model. The Appendices~\ref{app:microscopic} and \ref{app:susy} contain supporting material for the discussion in Section~\ref{sct:scft}. 

\section{The model} \label{sct:model}

We begin by reviewing the construction of the supersymmetric model of Refs.~\onlinecite{fendley2003,fendley2003-1} and discuss some important properties of the model.

To construct an $\mathcal{N}=2$ supersymmetric model,\footnote{In supersymmetric theories $\mathcal{N}$ refers to the number of supercharges. More precisely, an $\mathcal{N}=N$ supersymmetric quantum mechanical theory can be defined in terms of $N$ self-adjoint supercharges $Q_i=Q_i^{\dagger}$, where $i=1, \dots, N$, and a Hamiltonian, $H$. These operators act on a Hilbert space, $\mathcal{H}$, and obey the following anti-commutation relation for all $i,j=1, \dots, N$: $\{Q_i,Q_j\}=H\delta_{ij}$. Consequently, the Hamiltonian commutes with the supercharges and can be written as $H=\frac{2}{N} \sum_i Q_i^2$. For $\mathcal{N}=2$ supersymmetry it is more common, however, to define the theory in terms of the supercharge, $Q=(Q_1+\imath Q_2)\sqrt{2}$, and its adjoint, $Q^{\dagger}=(Q_1+\imath Q_2)\sqrt{2}$. The Hamiltonian then reads $H=\{Q,Q^{\dagger}\}$. This is the notation we use here.} we provide two nilpotent operators, referred to as supercharges and denoted as $Q$ and $Q^\dagger$. In our example, we construct these from fermionic operators that create "hard-core fermions", i.e. fermions with an infinite nearest-neighbor repulsion. These can be written in terms of normal fermionic creation and annihilation operators on the lattice $c_i^\dagger$ and $c_i$ using a projection operator
\begin{equation}
P_i = \prod_j (1-c^\dagger_j c_j),
\end{equation}
where the $j$ run over nearest neighbors of site $i$. Note that $P_i$ commutes with $c_j$ and $c_j^\dagger$ if either $i=j$ or $i$ and $j$ are not nearest neighbors; furthermore, $P_i^2 = P_i$. The hard-core fermion operators are then given by
\begin{align}
d_i &= P_i c_i &d_i^\dagger = P_i c_i^\dagger.
\end{align}
We note that closely related models for hard-core fermions on the square ladder have been studied previously in Refs.~\onlinecite{cheong2009,cheong2009-1,muender2010}; however, these models were not tuned to be supersymmetric.

We define the supercharges
\begin{align}
Q^\dagger &= \sum_i c_i^\dagger P_i &
Q &= \sum_i c_i P_i.
\end{align}
It is easily checked that these are nilpotent, $Q^2 = (Q^\dagger)^2 = 0$, and hence $\lbrace Q,Q \rbrace = \lbrace Q^\dagger, Q^\dagger \rbrace = 0$.

The supersymmetric Hamiltonian is now constructed as $H=\lbrace Q^\dagger, Q \rbrace$. It follows immediately that $[H, Q] = [H, Q^\dagger ] = 0$, i.e. the supercharges are conserved quantities.

A short calculation shows that
\begin{align}\label{eqn:Ham}
H &= \sum_{\langle i,j \rangle} \left( P_i c_i^\dagger c_j P_j + P_j c_j^\dagger c_i P_i \right) + \sum_i P_i.
\end{align}
The first term is a hopping term dressed with the hard-core projection to ensure that only states in the allowed subspace are generated. The second term can be interpreted as a potential term whose precise structure depends on the lattice. The Hamiltonian preserves the number of fermions, $[H, \sum_i n_i] = 0$, where $n_i=c_i^\dagger c_i$. We identify states with an odd number of particles as "fermionic" states, and "bosonic" otherwise.

Supersymmetry strongly constrains the spectrum of the model. It is easily shown that all energy eigenvalues $E_n$ obey $E_n \geq 0$. In addition, the eigenvectors corresponding to non-zero eigenvalues can be grouped in pairs $(|\psi\rangle, Q |\psi\rangle)$, where the two states differ in occupation by one fermion. This is reminiscent of the symmetry between bosonic and fermionic excitations in supersymmetric field theories.

It is convenient to write the model on the square ladder in terms of operators $d_{i,\up}^\dagger$ ($d_{i,\dw}^\dagger$), which create a fermion on the upper (lower) site of the $i$'th rung, and the corresponding number operators $n_{i,\up} = d^\dagger_{i,\up} d_{i,\up}$ and $n_{i,\dw} = d^\dagger_{i,\dw} d_{i,\dw}$. In terms of these operators, the non-trivial interaction terms of the potential part of the Hamiltonian can be written as:
\begin{align}
H_{v2} &= 2 \sum_i \left( n_{i,\dw} n_{i+1,\up} + n_{i,\up} n_{i+1,\dw} \right) \label{eqn:Jterm} \\
H_{v2'} &= \sum_i \left( n_{i,\dw} n_{i+2,\dw} + n_{i,\up} n_{i+2,\up} \right)  \\
H_{v3} &= - \sum_i \left( n_{i,\dw} n_{i+1,\up} n_{i+2,\dw} +
n_{i,\up} n_{i+1,\dw} n_{i+2,\up} \right) \label{eqn:interaction}
\end{align}
where $H_{v2}$ is a next-nearest neighbor repulsion across the diagonals of the plaquettes, $H_{v2'}$ is a next-nearest neighbor repulsion on the same chain, and $H_{v3}$ is an attractive three-body term.

Many insights into the behavior of the system can be gained by studying the behavior under perturbations. It is often favorable to study perturbations that preserve the supersymmetry since the same analytical tools can be applied. A large class of such perturbations is the staggering of the supercharges. The latter can be modified to contain a site-dependent complex factor, i.e.
\begin{align}
Q^\dagger &= \sum_i \bar{y}_i c_i^\dagger P_i &Q &= \sum_i y_i c_i P_i
\end{align}
where the $y_i$ are complex numbers, and $\bar{y}_i$ indicates complex conjugation. None of the algebraic properties of the model are changed under this modification. In particular, the number of ground states remains unchanged for arbitrary staggering. The behavior of other one-dimensional realizations of the model under staggering was investigated in Refs.~\onlinecite{huijse2011-1,fendley2011}.

The ground state degeneracy on the square ladder depends on the boundary conditions and the length of the system. For open and periodic boundary conditions the number of zero energy states can be determined analytically.\cite{fendley2003,HuijseThesis} For periodic boundary conditions there are 3 zero energy states if the length is a multiple of 4, and 1 otherwise. For open boundary conditions there is a unique zero energy state both for even and odd length of the ladder. For antiperiodic boundary conditions, the ground state is unique when the length is a multiple of 4. In all cases the ground state is at quarter filling.

\section{The second $\mathcal{N}=2$ supersymmetric minimal model} \label{sct:scft}

It was first conjectured in Ref. \onlinecite{fendley2003} that the continuum theory that describes the supersymmetric model on the two-leg ladder is the second model in the series of superconformal minimal models with $\mathcal{N}=2$ supersymmetry. Further support for this conjecture was presented in Refs.~\onlinecite{huijse2008,vdNoort,HuijseThesis}. Let us briefly comment on the continuum model, before we give a short review of these ideas and result.

Conformal field theories are characterized most importantly by their central charge $c$ and the scaling dimensions $h$. Without supersymmetry, the unitary rational CFTs with $c < 1$ form a discrete set referred to as the minimal models.~\cite{friedan1984} It can be enumerated by an integer $m \geq 3$; the central charge and scaling dimensions are given by
\begin{subequations} \label{eqn:minimal1} \begin{align}
c &= 1-\frac{6}{m(m+1)} \\
h &= \frac{ [r(m+1)-sm]^2 - 1}{4m(m+1)}
\end{align} \end{subequations}
where $r$, $s$ are integers with $1 \leq r \leq m-1$ and $1 \leq s \leq m$. For example, the two-dimensional classical Ising model at criticality corresponds to the first minimal model $m=3$.

With $\mathcal{N}=2$ supersymmetry, a similar list of minimal models can be given for central charges $1 \leq c < 3$.\cite{qiu1987} The minimal series in the case of $\mathcal{N}=2$ supersymmetry is also enumerated by an integer $k \geq 1$ and has
\begin{subequations} \label{eqn:minimal2} \begin{align}
c &= 3 - \frac{6}{k+2} \\
h &= \frac{p(p+2) - r(r-2) - 4r\alpha + 2k (\frac{1}{2} - \alpha)^2}{4(k+2)}
\end{align} \end{subequations}
for $0 \leq p \leq k$ and $r = -p, -p+2, \ldots, p$, and $\alpha = 0$ ($\alpha=1/2$) for the Ramond (Neveu-Schwarz) sector. In the Ramond and Neveu-Schwarz sectors the fermionic fields obey anti-periodic and periodic boundary conditions on the plane, respectively. The converse is true for the cylinder, where the Ramond and Neveu-Schwarz sector correspond to periodic and anti-periodic boundary conditions in the lattice model, respectively. A characteristic feature of supersymmetric theories is the Witten index, which is related to the number of Ramond vacua or zero energy states. The Witten index for the $k$-th minimal model is $W=k+1$.

The theory that has been conjectured to describe the continuum theory of the supersymmetric model on the square ladder is the $k=2$ supersymmetric minimal model with $c=3/2$. An overview of other possible SCFTs with $c=3/2$ is given in Ref.~\onlinecite{dixon1988}. Physically, this minimal model can be understood as the product of a compactified boson and an Ising theory (free Majorana fermion).
The lagrangian density can be written as
\begin{eqnarray}\label{eqn:Lscft}
\mathcal{L}&=&\mathcal{L_B}+\mathcal{L_F}, \nonumber\\
\mathcal{L_B}&=&\frac{1}{2 \pi} \partial \Phi \bar{\partial} \Phi, \\
\mathcal{L_F}&=& 2 (\psi_{R}\partial \psi_{R}+\psi_{L}\bar{\partial}\psi_{L}) . \nonumber 
\end{eqnarray}
The primary operators and their scaling dimensions in the Ising sector are
\begin{subequations} \begin{align}
\sigma_L, h_\sigma &= 1/16 \\
\psi_L, h_\psi &= 1/2
\end{align} \end{subequations}
for the left-movers and similarly for the right-movers. From the left- and right-moving bosonic fields, $\Phi_{L,R}$, we form vertex operators
\begin{subequations} \begin{align}
V_{m,n} &= e^{i (m+n) \Phi_L / \sqrt{2} +i (m-n) \Phi_R / \sqrt{2} } \\
h_{L,R} &= (m\pm n)^2/4
\end{align} \end{subequations}
where we have fixed the compactification radius at $r=\sqrt{2}$. The labels $m$ and $n$ are related to momentum and charge, respectively. In Appendix~\ref{app:microscopic} we discuss a way to relate the operators in the field theory to operators in the lattice model. The full operator content follows from considering products of the Ising and boson operators. The relevant operators are listed in Table~\ref{fig:relops}. Note that the model is tuned to the Kosterlitz-Thouless (KT) transition, where the cosine term, $\cos(2\Phi/\sqrt{2})$, is precisely marginal ($h_L+h_R=2$).

\begin{table}
\centering
\begin{tabular}{|c|c|c|c|c|} \hline
Operator &$h_L+h_R$ &SUSY \\ \hline \hline
$1$ &$0$ &Yes \\ \hline
$V_{\pm1,0}$ &$1/2$ &No \\ \hline
$\psi_L \psi_R$ &$1$ &No \\ \hline
$\psi_L \psi_R V_{\pm 1,0}$ &$3/2$ &Yes \\ \hline
$\sigma_L \sigma_R V_{\pm 1/2,0}$ &$1/4$ &No \\ \hline
$\sigma_L \sigma_R V_{\pm 3/2,0}$ &$5/4$ &Yes \\ \hline
\end{tabular}

\caption{Relevant operators of the continuum theory. The second column lists their respective dimension, and the third column indicates whether they preserve supersymmetry. \label{fig:relops}}
\end{table}

To see that this model is supersymmetric, remember that the left (right) moving supercharges have scaling dimension $(h_L,h_R)=(3/2,0)$ ($(h_L,h_R)=(0,3/2)$). It follows that any $c=3/2$ theory contains a left- and right-moving supercharge given by $\psi_L \partial \Phi_L$ and $\psi_R \bar{\partial} \Phi_R$. These supercharges generate an $\mathcal{N}=(1,1)$ supersymmetry, that is, an $\mathcal{N}=1$ supersymmetry in both the left- and the right-moving sector (remember that $\mathcal{N}$ stands for the number of supercharges). For the compactification radius $r=\sqrt{2}$ there is an additional $\mathcal{N}=(2,2)$ supersymmetry, generated by the two left- and right-moving supercharges $\psi_{L,R} \exp[\pm \imath \sqrt{2} \Phi_{R,L}]$.

We now return to the lattice model and briefly review the argument that led to propose the $c=3/2$ superconformal field theory as its continuum theory and some results that are in agreement with this conjecture. We first note that the Witten index, $W=k+1$, is equal to 3 for the 2nd minimal model. This is in agreement with the Witten index of the lattice model, given by $W=\textrm{Tr} (-1)^{\sum_i n_i}$, which is 3 if the length is a multiple of 4.~\cite{fendley2003} Furthermore, the microscopic Hamiltonian, Eq. \ref{eqn:Ham}, describes a
one-dimensional system of particles with strong repulsive
interactions. Due to the infinite nearest-neighbor repulsion,
there can be at most one particle per rung of the ladder;
therefore, one can describe the system in terms of a
``charge'' (longitudinal) degree of freedom, corresponding to the
position of the particles along the ladder, and an internal
``spin'' variable denoting the leg index of each particle. The
symmetry of the system under reflection, which maps the upper leg to the lower leg and vice versa,
implies that the spin degree of freedom has Ising-like ($Z_2$)
symmetry.
As was proposed in Ref.~\onlinecite{fendley2003}, it is suggestive to associate the spin variable with the Ising sector in the $c=3/2$ theory. The $U(1)$ symmetry associated to the charge degrees of freedom then simply corresponds to the bosonic sector. Further evidence for this picture was given in Ref.~\onlinecite{vdNoort}, where two exactly soluble limits of this model were considered by perturbing away from the supersymmetric point. It was shown that in these limits, the model exhibits an Ising transition and a KT transition, respectively. It was suggested that these two transitions coincide at the supersymmetric point. We briefly review the soluble limits in App.~\ref{app:sollimits}. Finally, it was shown in Ref.~\onlinecite{huijse2008} that the supersymmetric model is gapless by numerical studies of a boundary twist using exact diagonalization. These numerical studies, however, also revealed the strong finite-size effects that made it thus far impossible to conclusively identify the $c=3/2$ theory as the continuum theory of this model.~\cite{HuijseThesis}


\subsection{Marginal operators}\label{sct:marginal}

The operator content of our candidate theory in principle allows for a variety of marginal operators, i.e. operators with scaling dimension $h=h_L+h_R=2$. These could serve as an explanation of the strong finite-size effects observed in numerical calculations. On the other hand, one might naively expect that these operators are excluded by the explicit supersymmetry on the lattice. We have found, however, that there is a special combination of two marginal operators that preserves $\mathcal{N}=2$ supersymmetry (see Appendix~\ref{app:susy}). The two marginal operators are the usual cosine term for the boson, $\cos(\sqrt{2} \Phi)$, and the more unusual operator $(\partial_x \Phi) \imath \psi_R \psi_L$. In the following we
analyze the renormalization group flows of the $c=3/2$ theory in
the presence of these operators without imposing supersymmetry. At
the end of this section we will discuss the flow along the
direction that preserves supersymmetry to see the finite-size
effects resulting from the presence of these marginal operators.

The renormalization group flow of the $c=3/2$ theory in the
presence of the operator $(\partial_x \Phi) \imath \psi_R \psi_L$
was worked out in Ref.~\onlinecite{Sitte} by considering Wilsonian
one-loop momentum-shell RG. In order to analyze the cosine term it
will be more convenient to work in real-space and use the Operator
Product Expansion formulation of RG.~\cite{Cardy_Scaling_RG} 
We consider the lagrangian density
$\mathcal{L}=\mathcal{L_B}+\mathcal{L_F}+\mathcal{L_{\rm{int}}}$,
where
\begin{eqnarray}
\mathcal{L_B}&=&\frac{1}{2 \pi K} \Big( \frac{1}{v}(\partial_{\tau}\Phi)^2 + v(\partial_{x}\Phi)^2 \Big) \nonumber\\
\mathcal{L_F}&=&\psi_{R}(\partial_{\tau}+\frac{u}{\imath}\partial_{x})\psi_{R}+\psi_{L}(\partial_{\tau}-\frac{u}{\imath}\partial_{x})\psi_{L} \nonumber\\
\mathcal{L_{\rm{int}}}&=& -\lambda (\partial_x \Phi) \imath \psi_R
\psi_L + g \cos[\sqrt{2} \Phi]. \label{eqn:L}
\end{eqnarray}
Here, $v,K$ are the velocity and Luttinger parameter of the
bosonic field, $u$ is the fermion velocity, and $\lambda,g$ are
coupling constants. Note that the lagrangian density of the superconformal field theory (\ref{eqn:Lscft}) is obtained for $u=v$, $K=4$ and $\lambda=g$. With these conventions the scaling
dimension of the cosine term is $h= K/2$, which is marginal at
$K=4$. We treat the interaction terms perturbatively. In a
real-space RG approach one takes care of UV divergencies by
introducing a real-space cutoff, $a$, which defines the minimal
distance that two operators can approach each other. In an RG step
this cutoff is increased, $a \to (1+\delta \ell)a$, while the effective action is kept fixed by
renormalizing the couplings. We work in a scheme where we keep the
bosonic velocity $v$, the normalization of the field $\Phi$ and
the unit prefactor of the term $\sum_{i=L,R}
\psi_{i}\partial_{\tau}\psi_{i}$ fixed. To achieve this we allow
the fermion field to rescale, $\psi \to \sqrt{Z_F} \psi$, and we
introduce an anomalous dynamical exponent, $z$. Having $z\neq 1$
reflects the fact that the $\lambda$-term breaks Lorentz
invariance. For the field rescaling we find
\begin{eqnarray}
Z_{F}&=&1+\frac{\lambda^{2}K}{8(v+u)^{2}},\nonumber
\end{eqnarray}
and for the dynamical exponent we obtain
\begin{eqnarray}
z&=&1+\frac{\lambda^{2}K}{16 uv}.\nonumber
\end{eqnarray}
The fermionic velocity, $u$, the Luttinger parameter, $K$, and the
interactions $\lambda$ and $g$ flow under RG. To second order in
$\lambda$ and $g$ we obtain the following RG equations
\begin{eqnarray}
\frac{du}{d\ell}&=&-\frac{u\lambda^{2}K}{4}\Big(\frac{1}{(v+u)^{2}}-\frac{1}{4uv}\Big)\nonumber\\
\frac{d\lambda}{d\ell}&=&0\nonumber\\
\frac{dK}{d\ell}&=&-K^{2}\Big(\frac{g^{2}\pi^2}{2v^{2}}-\frac{\lambda^{2}}{16 uv}\Big)\nonumber\\
\frac{dg}{d\ell}&=&\frac{g}{2}(4-K).\nonumber
\end{eqnarray}
The first two equations and the dynamical exponent precisely agree
with the results of Ref.~\onlinecite{Sitte} (up to slight differences in conventions\footnote{Note that the
conventions differ slightly. To compare directly with
Ref.~\onlinecite{Sitte} one has to take $\lambda \to 2
\lambda/\pi$.}). The latter two equations reduce to the Kosterlitz
equations, when we set $\lambda$ to zero. Indeed, for $\lambda=0$
the boson and Ising sectors are completely decoupled. Finally, we
find that for $u=v$ and $K=4$ there is a line given by
\begin{eqnarray}
\lambda=2\sqrt{2}\pi g,\nonumber
\end{eqnarray}
where the RG equations are all zero to second order in the
couplings. Note that for this value of $K$ the cosine term is
indeed marginal. As shown in Appendix~\ref{app:susy}, precisely this line
preserves $\mathcal{N}=2$ supersymmetry. 
It follows that coupling to this special combination of these two marginal operators is not
excluded by the explicit lattice supersymmetry.

The line of fixed points is likely to disappear at higher order in
perturbation theory.\cite{Sitte,HuijseBergBauer} 
Based on our knowledge of our lattice model, we argue 
that to higher order there will be a flow towards the fixed point
$\lambda=g=0$: (i) The
central charge of the $\mathcal{N}=2$ superconformal minimal
models is given by $c=3k/(k+2)=1,3/2,9/5,2,\dots$. Our numerical
analysis of the central charge gives a value close enough to
$c=3/2$ to be able to exclude the other minimal models (see Sec.~\ref{sct:resA}).
(ii) Using the fact that for $L=4n$ the lattice model has 3 zero energy
states in the Ramond sector (pbc) and 1 negative energy state in
the NS sector (apbc), we can exclude other $c=3/2$ theories with
higher supersymmetry.~\cite{HuijseThesis}

The extremely slow flow along the supersymmetric line parametrized by
$\lambda$ leads to strong finite-size corrections. In particular, we expect slowly decaying corrections to the energy and effective central charge. Since there is no flow to second order, we expect that $d\lambda/d\ell = B\lambda^3 + O(\lambda^4)$. The
explicit computation of $B$ will be presented
elsewhere.\cite{HuijseBergBauer} For now, we will assume that
$B<0$, consistent with the DMRG results that indicate that the
$c=3/2$ fixed point is stable in the presence of supersymmetry.
Solving the flow equation for $\lambda$, we then find
$\lambda(\ell) \sim 1/\sqrt{\ell}$. For the scaling of energy
levels this leads to a finite size scaling of the form,~\cite{cardy1986}
\begin{equation}
E(L)=E/L+\alpha_1/(L \sqrt{\log(L)})+\dots,
\end{equation}
where $\alpha_1$ is some non-universal parameter. 
Clearly, one has to go to extremely large system
sizes to reliably extract the universal value $E=h_L+h_R-c/12$.
For the central charge, however, it turns out that the finite size
corrections are strongly suppressed. The scaling form of the
central charge due to the presence of a marginally irrelevant
operator is \cite{cardy1986,affleck1989}
\begin{eqnarray}\label{eqn:effc}
c(L)&=& c+\alpha_2 \lambda(L)^3 + \dots \nonumber\\
&=&c+\alpha  \frac{1}{(\sqrt{\log(L)} )^3}+ \dots
\end{eqnarray}
where $\alpha$ is a fit parameter. We will discuss the implications of this effective central charge for our numerical approach in the Sections~\ref{sct:identification} and \ref{sct:resA}.

\section{Methods} \label{sct:methods}

\subsection{DMRG}

Our calculations are carried out using the density matrix renormalization group method.~\cite{white1992,white1992-1,schollwoeck2005,schollwoeck2011} This method is generally formulated for Hilbert spaces with a tensor-product structure and therefore does not allow the implementation of the hard-core constraint on the level of the Hilbert space. Instead, we add a penalty term to the Hamiltonian that increases the energy of configurations with occupied nearest-neighbor sites. We find that this term does not have to be very strong to obtain reliable convergence as the hopping term is dressed with the projection operator and therefore only acts on the allowed subspace. In addition, we choose a basis where the two sites on a rung are treated as a single site, allowing us to implement the hard-core constraint exactly for this rung.

The approximation made in DMRG calculations can be systematically refined by increasing the number of states $M$ kept in the renormalization procedure. The algorithmic cost grows as $\mathcal{O}(M^3)$, which limits the number of states to a few thousand in practical calculations. While for gapped systems, a bond dimension of a few hundred is generally sufficient independent of system size, critical systems are more challenging because the number of states has to be increased as some polynomial of the system size when the thermodynamic limit is approached. In addition, periodic boundary conditions strongly increase the number of states needed when using a standard DMRG approach. While improved schemes exist,~\cite{schollwoeck2005,schollwoeck2011} we use a well-tested and numerically robust standard approach. While most of our simulations are carried out using $M=1000$ states, we confirm results for up to $M=4800$ states for some long periodic systems. This limits us to systems of length up to $L=100$.

\subsection{Identification of the conformal field theory} \label{sct:identification}

Correctly identifying the field theory that describes the continuum limit of a given microscopic model is a notoriously difficult problem. In general, one has to resort to numerical simulations, which are usually restricted to finite systems. While the most commonly used method of identifying the CFT is by fitting the spectrum obtained with exact diagonalization for small systems to a spectrum obtained directly for the CFT, we use a different approach in this paper: we first establish the central charge at the supersymmetric point by studying the entanglement entropy of the system (Sct.~\ref{sct:resA}). We then move on to study the phase diagram of the model with various supersymmetry-breaking perturbations, where we again rely on entanglement entropy -- augmented with calculations of correlation functions or structure factors -- to establish the phases (Sct.~\ref{sct:pd}). Finally, we study a perturbation where in one limit, the model in the exactly soluble case of a chain is recovered (Sct.~\ref{sct:zstag}).

As discussed above, finite-size corrections to the central charge due to marginal operators are strongly suppressed compared to corrections in other quantities. Calculating the central charge therefore appears as a much more viable approach to identifying the CFT than studying the energy spectrum. We will now elaborate on how this can be done in numerical calculations, and then discuss possible corrections in finite systems.

We consider the entanglement entropy $S(\rho) = -\text{Tr}\ \rho \log \rho$ where $\rho$ is the reduced density matrix for some block of sites.
If we denote the entanglement entropy between a block of $l$ contiguous sites in an infinite system and the rest of the system as $S(l)$, we have for the entropy of this block~\cite{holzhey1994}
\begin{equation}
S(l) \sim \frac{c}{3} \log l.
\end{equation}
Using this relation, the central charge can be extracted by calculating the entanglement entropy at the center of a finite system of length $L$ and performing a fit to
\begin{equation}
S(l=L/2) \sim \frac{nc}{6} \log L, \label{eqn:S_simple}
\end{equation}
where $n=1$ for open and $n=2$ for periodic boundary conditions. For reasonably large systems, which can be simulated with the DMRG method, such a fit often gives accurate results for the central charge.

For a finite block embedded in a finite systems, Eqn.~\eqnref{eqn:S_simple} only holds approximately. A more appropriate relation for finite systems is obtained in Ref.~\onlinecite{calabrese2004}. The entropy for a block of $l$ sites in a finite periodic system of length $L$ is
\begin{equation}
S(l) = \frac{c}{3} \log \left( \frac{L}{\pi} \sin \frac{\pi l}{L} \right) + S_0 \label{eqn:S_pbc},
\end{equation}
and for $l$ sites at the end of an open system we have
\begin{equation}
S(l) = \frac{c}{6} \log \left( \frac{2L}{\pi} \sin \frac{\pi l}{L} \right) + S_0 \label{eqn:S_obc}.
\end{equation}
In the constants $S_0$, we have summed up several universal and non-universal contributions, which are not relevant for our purposes. This expression opens up a second way of determining the central charge: instead of simulating several different system sizes and performing a fit to Eqn.~\eqnref{eqn:S_simple}, one can simulate only a single system size and calculate the entanglement entropy for various block sizes $l$. A fit to Eqns.~\eqnref{eqn:S_pbc} or \eqnref{eqn:S_obc} will then yield the desired estimate for the central charge.

It is important to note that when performed in a finite system, both of the procedures described above only obtain an estimate for the central charge and not the exact value. In the remainder of this paper, we will denote the estimate for the central charge obtained from such a fit as $\ceff$ to avoid any confusion. If the system is described by a conformal field theory in the thermodynamic limit, this estimate will approach the true central charge of that CFT as the system size is increased. In the case considered in this paper, we can quantify the leading finite-size corrections that occur in this approach and show that the data agrees with the prediction, cf. Section~\ref{sct:marginal}.

An important situation to consider is that of a gapped system which is very close to a critical point and hence has a very large correlation length $\xi$. If the above fitting procedures are performed for system sizes comparable to or smaller than the correlation length, a reasonable fit may be obtained for Eqns.~\eqnref{eqn:S_pbc},~\eqnref{eqn:S_obc} with a fit coefficient $\ceff$ that is related to the properties of the nearby critical point. However, this is purely an artifact of the insufficient system size and the fit procedure will break down as the system size is increased to be sufficiently large compared to the correlation length. In practice, one will observe $\ceff \rightarrow 0$ for $L \rightarrow \infty$ (and in particular $L \gg \xi$) in such a case since $S(l)$ becomes independent of $l$ for $l \gg \xi$. It is an important but unresolved question whether a scaling form for the entanglement entropy exists in the vicinity of a critical point.

Calculating $\ceff$ in the vicinity of a critical point which can be tuned by some parameter $t$, one would observe the following behavior: at the location of the critical point, $t = t_c$, $\ceff \rightarrow c$ for $L \rightarrow \infty$, where $c$ is the central charge of the CFT that governs this critical point. For $t \neq t_c$, one will observe $\ceff \rightarrow 0$ for $L \rightarrow \infty$, where the limit is approached more quickly the smaller the correlation length is, i.e. the further away the system is tuned from the critical point. Calculating $\ceff$ for a range of parameters around the critical point, one will therefore observe a peak which sharpens as the system size is increased, and which will ultimately approach $\ceff = c \delta(t-t_c)$. Very similar behavior occurs when a transition occurs between two critical phases; in such a case, if the central charges of the adjacent critical phases are $c_1$ ($c_2$), and the transition between them has $c_3 > c_1$ and $c_3 > c_2$, one would observe $\ceff \rightarrow c_1$ ($\ceff \rightarrow c_2$) as $L \rightarrow \infty$ within the respective phases, and $\ceff \rightarrow c_3$ at the transition. Again, finite-size corrections would be more pronounced close to the transition, leading to a finite-width peak if $\ceff$ is measured in a finite system.
We will make use of this approach later to map out the phase diagram of our model. We point out that this is very similar to approaches using entanglement measures directly to probe quantum phase transitions.~\cite{vidal2003,osterloh2002,osborne2011}


We also note that additional corrections occur for open boundary conditions. These were first observed numerically in Ref.~\onlinecite{laflorencie2006} and later explained analytically in Refs.~\onlinecite{cardy2010,fagotti2011}. The correction can be attributed to Friedel oscillations from the boundary, and is well fit by
\begin{equation}
S^c(l) \sim \left( \frac{L}{\pi} \sin \frac{\pi l}{L} \right)^{-K},
\end{equation}
where $K$ is the Luttinger liquid parameter in the case of $c=1$ CFTs, and related to the scaling dimension of relevant operators otherwise. The correction can heuristically be explained by a small dimerization on open lattices. Due to this correction, we will restrict our calculations to periodic systems.

\section{Results} \label{sct:results}

\subsection{The SUSY point}\label{sct:resA}

\begin{figure}
  \includegraphics{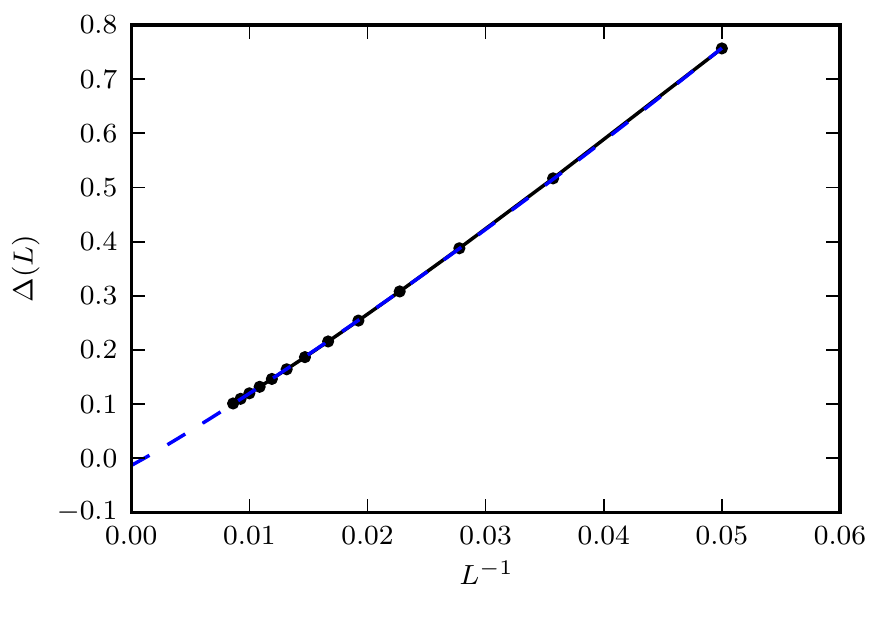}
  \caption{Finite-system gap for the ladder with open boundary conditions. Calculations were performed with $M=1000$ states. The dashed line shows a fit to Eqn.~\eqnref{eqn:gap}. \label{fig:gap} }
\end{figure}

We start by demonstrating that the system is indeed gapless with a dynamical critical exponent of $z=1$, i.e. that $\Delta(L) \sim \xi^{-1}$, and since the correlation length is expected to diverge at a critical point, $\Delta(L) \sim L^{-1}$ (to leading order). This is a the necessary condition for the system to be described by a conformal field theory in the continuum limit.

In Fig.~\ref{fig:gap}, the finite-size charge gap, defined as $\Delta(L) = E(N=L/2+1)+E(N=L/2-1)$, where $N$ is the number of particles, is shown. Note that the ground state has $N=L/2$, and $E(N=L/2)=0$ by supersymmetry. Since this calculation is performed for open boundary conditions, a modest number of states in the DMRG procedure is sufficient to obtain good accuracy even for systems of length $L > 100$. We perform an extrapolation by a fit to
\begin{equation} \label{eqn:gap}
\Delta(L) = \Delta_0 + \frac{\alpha_1}{L} + \frac{\alpha_2}{L \sqrt{\log L}}.
\end{equation}
We expect in particular that $\Delta_0 = 0$, i.e. the system is gapless. The subleading terms are included due to the presence of marginal operators in the candidate CFT (see Sec.~\ref{sct:marginal}). Such correction terms were discussed in Refs.~\onlinecite{cardy1986,affleck1989}. The good agreement with the expected scaling confirms that the system is indeed described by a conformal field theory perturbed by marginal operators. The accuracy could be improved by including further higher-order corrections.

\begin{figure}
  \includegraphics{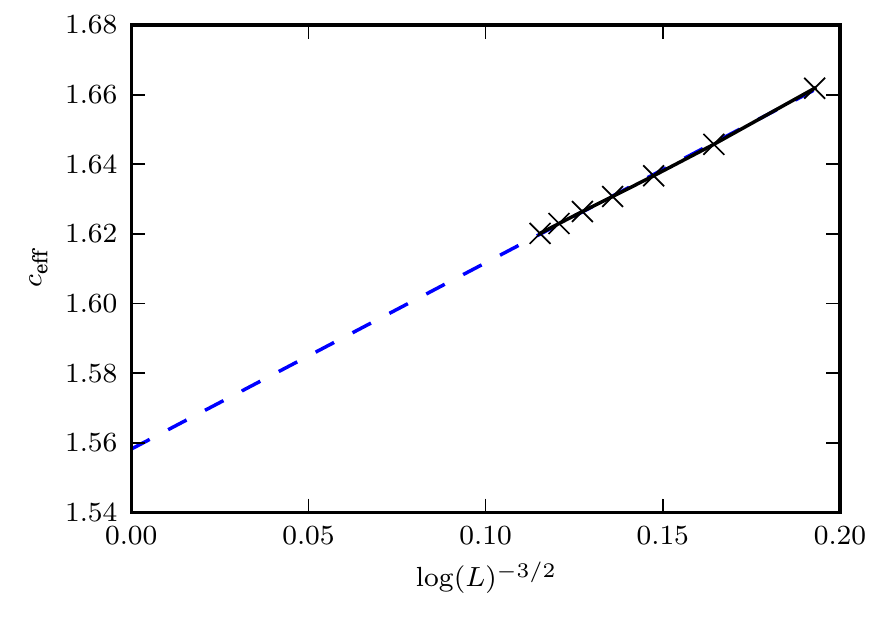}
  \caption{Effective central charge obtained for a ladder with antiperiodic boundary conditions using $M=4800$ states. \label{fig:ladder_direct} }
\end{figure}

We next move on to study the entanglement entropy at the supersymmetric point. In order to select a unique ground state and avoid oscillating terms from the boundaries, we use antiperiodic boundary conditions for this calculation. Due to this choice of boundary conditions, we have to use a large number of states in a conventional DMRG procedure in order to obtain accurate results also for large systems. In Fig.~\ref{fig:ladder_direct}, data is shown that was obtained using $M=4800$ states, which is sufficient to exhaust the entropy for these system sizes. The effective central charge we obtain shows large finite-size corrections, which is unusual. We will now show that these corrections can consistently be explained by the presence of marginal operators.

To account for finite-size corrections due to marginal operators, we perform an extrapolation of the effective central charge $\ceff$ measured in a finite system of length $L$ using Eqn.~\eqnref{eqn:effc}, where we use both $c$ and $\alpha$ as fit parameters. We find good agreement and an extrapolated value reasonably close to the expected value $c=1.5$. This gives strong indication that the central charge is indeed $c=3/2$. It also shows that the corrections to the central charge originally obtained for the free energy density can analogously be applied to calculations based on the entanglement entropy.

\subsection{Supersymmetry-breaking perturbations} \label{sct:pd}

Our candidate $c=3/2$ CFT can be characterized as the product of a free boson (Luttinger liquid with $c=1$) and a critical Ising model ($c=1/2$). Furthermore, the bosonic part of the theory is tuned to the Kosterlitz-Thouless (KT) transition where the staggering operator is exactly marginal. Such a critical point is expected to appear as a multicritical point in a two-parameter phase diagram. One would expect four phases adjacent to this critical point, corresponding to ordered and disordered phases of the charge and the Ising degree of freedom. In this section we explore this two-parameter phase diagram by studying two lattice perturbations, introduced below, that take us away from the supersymmetric point.

Our system has charge conservation, providing an obvious identification for the free boson part of the theory. Since we consider spinless fermions, the Ising part of the theory cannot be related to particle spin. As was proposed in Sec.~\ref{sct:scft}, we will see that it is instead related to the lattice parity symmetry of exchanging the upper and lower chains.
We find that the phase diagram can most easily be obtained by tuning the following three parameters: i) the rung hopping $t_\perp$, which favors states odd under exchange of the sites on a rung, ii) a two-body density repulsion $J$ (defined below), which favors charge ordering, and iii) the filling. While in general only two parameters should need to be tuned to obtain such a phase diagram, we have found that one of the four possible phases adjacent to the supersymmetric point cannot be realized without changing the filling.

The perturbations can be written using the operators introduced in Section~\ref{sct:model} in the following form:
\begin{eqnarray}\label{eqn:pert}
H_\text{pert} &=& t_\perp \sum_i \left( d_{i,\dw}^\dagger d_{i,\up} + d_{i,\up}^\dagger d_{i,\dw} \right) \nonumber \\
&& + J \sum_i \left( n_{i,\dw} n_{i+1,\up} + n_{i,\up} n_{i+1,\dw} \right)
\end{eqnarray}
Note that in these units, $t_\perp = J = 0$ corresponds to the supersymmetric point. At $t_\perp = -1$, there is no rung hopping because the perturbing term exactly cancels the one in the original Hamiltonian. The term proportional to $J$ corresponds to the potential term $H_{v2}$ of Eqn.~\eqnref{eqn:Jterm}; thus, at $J=-2$, this term is removed from the Hamiltonian entirely. Unless otherwise mentioned, we will assume that the chemical potential has been adjusted such that the ground state remains at quarter filling. The filling is enforced exactly by using quantum numbers in the DMRG calculations. In some phases, we tune the filling away from the ground state filling $N = L/2$ by inserting a small number of holes.

Perturbations of this model away from the supersymmetric point were also considered in Ref.~\onlinecite{vdNoort}. It was found that in the limit $t_\perp=-1$ the model can be solved exactly by mapping it to an XXZ Heisenberg chain. The KT transition between the charge ordered and charge disordered phase is found to be at $J=0$ in this limit. Furthermore, in the limit $J \to \infty$ the model maps to an effective Ising model, with an Ising transition at $t_\perp=-1/2$. We give more details on these soluble limits in App.~\ref{app:sollimits}.

\begin{figure}
  \includegraphics[width=\columnwidth]{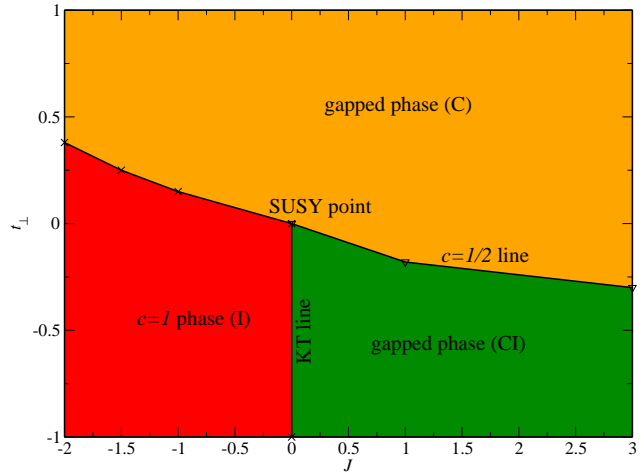}
  \caption{Phase diagram for the model in the $t_\perp$-$J$ plane. The labels show the shorthand notation for the phases: C corresponds to charge order, I to Ising order, and CI to charge and Ising order. 
\label{fig:pd} }
\end{figure}

\begin{figure}
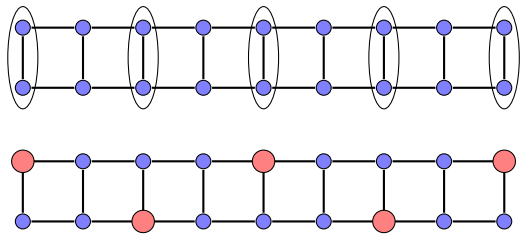

  \beginpgfgraphicnamed{fig_phases1}
    \input{phases1}
  \endpgfgraphicnamed
  \vspace{0.5cm}
  \beginpgfgraphicnamed{fig_phases2}
    \input{phases2}
  \endpgfgraphicnamed
  \caption{Illustration of the two gapped phases. Top panel: charge ordered, Ising disordered phase. The ellipses indicate sites occupied by a fermion in the antisymmetric state $(d_{i,\up}-d_{i,\dw})|0\rangle$. Bottom panel: charge and Ising ordered phase. The bigger red sites indicate occupied sites, the other sites are empty. \label{fig:gapped_phases} }
\end{figure}

We expect four possible phases adjacent to the supersymmetric point, which are characterized by charge (dis-)order and Ising (dis-)order. We find that for the case without doping away from quarter filling, only three phases are realized: (i) an Ising ordered, charge ordered phase which is fully gapped, (ii) a Ising disordered, charge ordered phase, which is also fully gapped, and (iii) an Ising ordered, charge disordered Luttinger liquid phase. We will refer to these phases using the shorthand notation C for charge ordered, I for Ising ordered, and CI for charge and Ising ordered. When the system is doped below quarter filling, the holes become itinerant in both of the fully gapped phases, giving rise to a gapless charge mode. This allows us to realize the (iv) Ising disordered, charge disordered phase. 

Phase C and CI are separated by a line of $c=1/2$ Ising transitions, and phase I and CI are separated by a line of KT transitions. At $t_\perp=-1$, the system can be solved exactly and the transition is found to be exactly at $J=0$ (see App.~\ref{app:sollimits}). The line of KT transitions and the line of Ising transitions join at the supersymmetric point. The nature of the transition from phase I to phase C is unclear and will be discussed below.


\subsubsection{Phase boundaries} \label{sct:resB1}

To establish the phase boundaries, we calculate the entanglement entropy for a fixed system size $L=24$ on a fine mesh of points in the $(t_\perp, J)$ plane. The phase boundaries are then easily extracted from the behavior of the effective central charge $\ceff$, as explained in detail in Section~\ref{sct:identification}. Our results are shown in Fig.~\ref{fig:pd}. As expected, the I phase appears as an extended region with central charge $c=1$. The precise location of the transition from this phase into the CI phase is known to lie at $J=0$ for $t_\perp=0$ and $t_\perp=-1$. For $-1 < t_\perp < 0$, the location is unknown, and due to the Kosterlitz-Thouless nature of the transition is difficult to determine numerically with a high accuracy. From our results for few system sizes, however, we can exclude significant deviations from the $J=0$ line. The transition from the CI to the C phase appears as a line of $c=1/2$ transitions, as expected. This is expected to approach $t_\perp=-0.5$ for $J \rightarrow \infty$, which is consistent with our observations. Note that the numerical results for the line of Ising transitions agrees well with the large J result, $t_{\perp}=-1/2+1/(J+2)$, even as $J \to 0$ (see App.~\ref{app:sollimits}).

The nature of the transition from the I phase to the C phase is at this point unknown. Several scenarios are possible: i) A line of $c=3/2$ transitions emanating from the supersymmetric point where the Ising transition and the charge ordering transition take place simultaneously. This requires fine-tuning and seems plausible only in the presence of an additional symmetry. ii) Without fine-tuning, the Ising transition could either be in the gapped or the charge-disordered phase. In the first case, it would correspond to a line of $c=1/2$ transitions, whereas in the latter case $c=3/2$ should be observed with $c=1$ on either side.

To obtain insights into this behavior, we calculate the entanglement entropy along several cuts of the phase diagram for fixed values of $J$. We then extract the effective central charge $\ceff$ as detailed in Section~\ref{sct:identification} as a function of $t_\perp$. Numerically, we find no evidence of two separate transitions. Instead, $\ceff$ shows a single, yet broad peak. This persists to system sizes up to $L \sim 70$ with antiperiodic boundary conditions, but the position of the peak shifts significantly with system size. This behavior seems more consistent with the first scenario of a single transition, where the peak would sharpen into a delta peak of $c=3/2$ at the transition from the gapless to the gapped phase if one could access sufficiently large system sizes. However, we observe an unusually large value of the effective central charge, which approaches $\ceff=2$ for $J \rightarrow -2$, which cannot be explained in this scenario. Additionally, since the observed peak is very broad, one could argue in favor of the second scenario where the two transitions are so close to each other that they cannot be resolved on the accessible system sizes. In conclusion, the precise nature of this transition remains an open question.

\subsubsection{Density structure factor} \label{sct:resB2}

We now discuss the exact ground states in two limits corresponding to the two gapped phases, C and CI, providing an intuitive picture for these phases. Furthermore, understanding these limits will allow us interpret the numerical results for the density structures in the different phases discussed below. In the C phase, we expect a state reminiscent of a valence-bond solid, where every other rung is occupied by an odd combination of the two states on the rung, i.e. the antibonding state. In the limit $t_\perp \rightarrow \infty$, the state should be a product state
\begin{equation} \label{eqn:gs_c}
|\psi\rangle = \left( \frac{1}{\sqrt{2}} \right)^{L/2} (d_{1,\dw}^\dagger - d_{1,\up}^\dagger) (d_{3,\dw}^\dagger - d_{3,\up}^\dagger) \ldots |0\rangle.
\end{equation}
This state is shown in the upper panel of Fig.~\ref{fig:gapped_phases}. In this limit, the translational symmetry is broken to a four-site unit cell and the state should be twofold degenerate. At finite values of $t_\perp$, we expect this degeneracy to be lifted since the symmetric and antisymmetric combination of the two translated product states are separated by an exponentially small energy gap.

The other fully gapped phase, CI, is characterized by particles alternating between the upper and lower chain with an empty rung between each particle. In the limit $t_\perp \rightarrow 0$, $J \rightarrow \infty$, the state is
\begin{equation} \label{eqn:gs_ci}
|\psi\rangle = d_{1,\dw}^\dagger d_{3,\up}^\dagger d_{5,\dw}^\dagger \ldots |0\rangle.
\end{equation}
This is shown in the lower panel of Fig.~\ref{fig:gapped_phases}.
Due to translational symmetry breaking, one would expect an four-fold degeneracy in the extreme limit, which is again lifted at finite values of $t_\perp$ and $J$.

The Ising ordered, charge disordered phase I can be understood by considering the CI phase discussed before, and allowing an overall charge mode. In particular, due to Ising order, the density correlations still show the alternating structure between the upper and lower chain. Similarly, one would expect an Ising disordered, charge disordered phase upon doping the C phase, where the antibonding states of phase (i) become itinerant. However, the infinite nearest-neighbor repulsion implies that two fermions cannot simultaneously be in the antibonding state on adjacent rungs. The Ising disordered, charge disordered phase can therefore not be realized using the parameters discussed so far, but instead must be realized by hole-doping the system away from quarter filling.

\begin{figure}
  \includegraphics{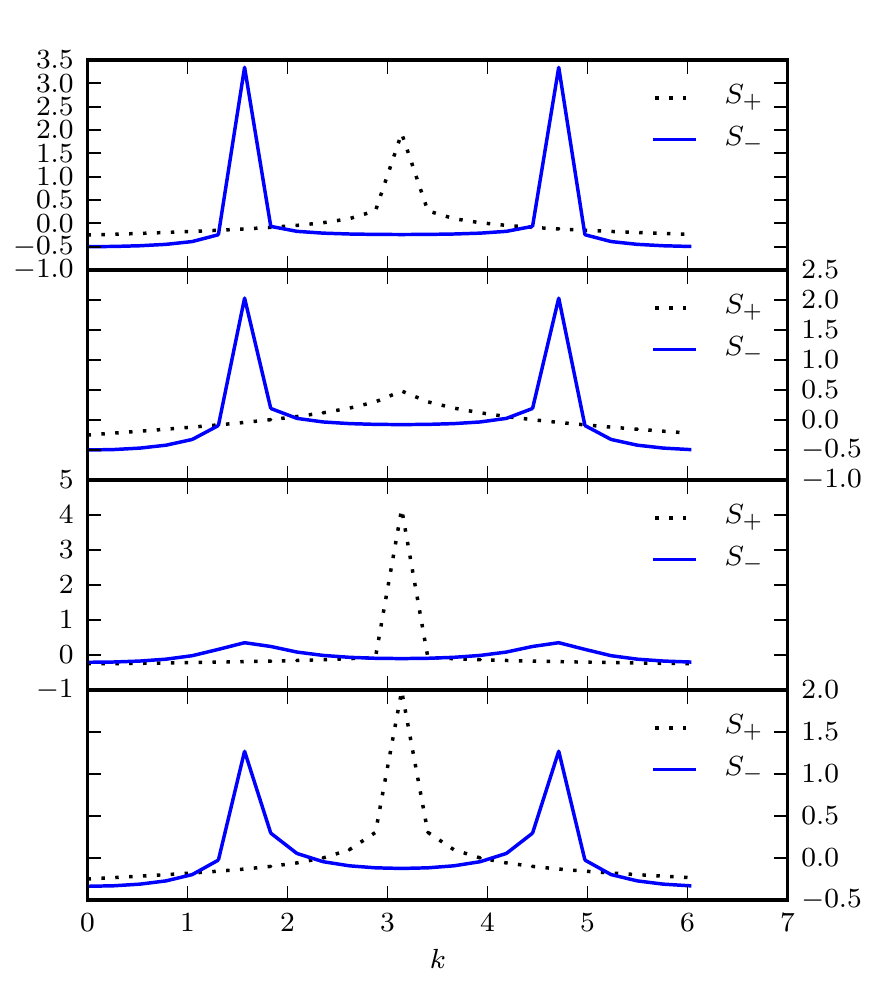}
  \caption{Structure factor for $L=24$ and antiperiodic boundary conditions. Top panel: Charge and Ising ordered phase, $J=1$, $t_\perp=-1$. Second panel: Charge disordered, Ising ordered phase, $J=-1$, $t_\perp=-1$. Third panel: Charge ordered, Ising disordered phase, $J=0$, $t_\perp = 0.65$. Bottom panel: Supersymmetric point. \label{fig:sf_1} }
\end{figure}


\begin{figure*}
  \begin{tabular}{cV}
  $J=1$, $t_\perp=-1$ &\includegraphics[width=5in]{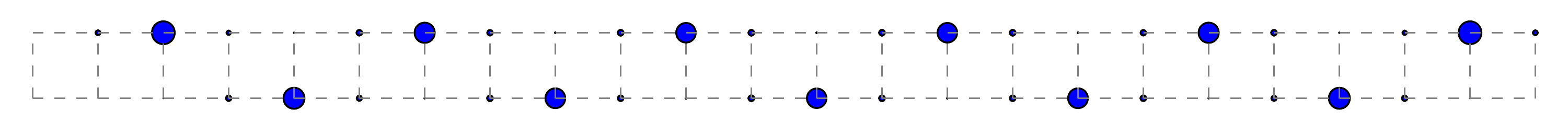} \\
  $J=-1$, $t_\perp=-1$ &\includegraphics[width=5in]{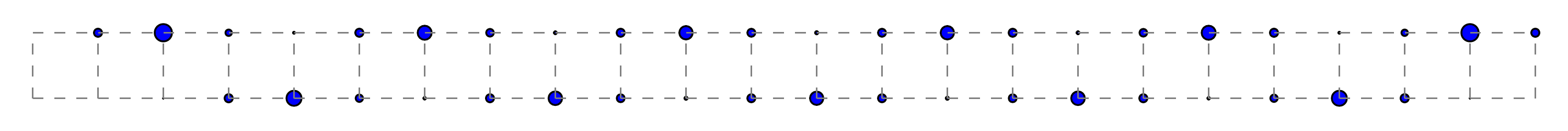} \\
  $J=0$, $t_\perp=0.65$ &\includegraphics[width=5in]{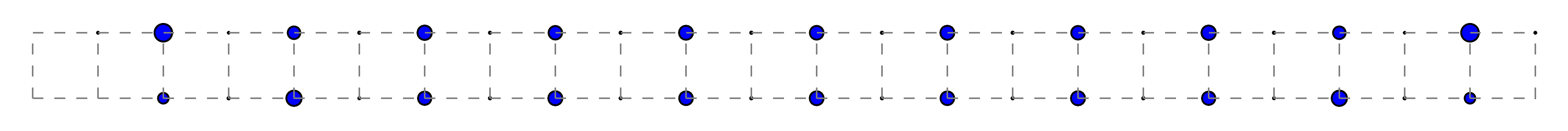} \\
  \end{tabular}
  \caption{Strength of the density correlation functions $\langle n_{0,\dw} n_{j,\up} \rangle$ and $\langle n_{0,\dw} n_{j,\dw}\rangle$ (shown only with $j > 0$) for values of the parameters representative of the three different phases at quarter filling. The diameters of the blue, filled dots are proportional to the strength of the correlation; the upper row in each panel corresponds to $\langle n_{0,\dw} n_{j,\up} \rangle$, while the lower row corresponds to $\langle n_{0,\dw} n_{j,\dw}\rangle$, and the distance from the left end indicates $j$. \label{fig:dens} }
\end{figure*}

We can now study the density structure factor to provide further evidence that the phases discussed above are present in the lattice model. Defining first the even and odd density on each rung,
\begin{align}
n_{i,+} &= n_{i,\up} + n_{i,\dw} & n_{i,-} &= n_{i,\up} - n_{i,\dw},
\end{align}
we can define the even and odd structure factors
\begin{equation} \label{eqn:sf}
S_\pm (k) = \sum e^{i k d} \left( \langle n_{i,\pm} n_{i+d,\pm} \rangle - \langle n_{i,\pm} \rangle \langle n_{i+d,\pm} \rangle \right)
\end{equation}
In a translationally invariant system, the index $i$ can be fixed to 0.

In Fig.~\ref{fig:sf_1}, the numerically calculated structure factor for a fixed system size and four different values of the parameters, corresponding to the three different phases and the supersymmetric point, are shown. By inspection, Ising order corresponds to peaks in the $S_-$ structure factor at $k=\pi/2$ and $k=3\pi/2$, whereas charge order corresponds to a peak in $S_+$ at $k=\pi$. To understand this behavior, one can consider the ground state in the limiting cases given in Eqns.~\eqnref{eqn:gs_c}, \eqnref{eqn:gs_ci}. In the C phase, $n_-$ vanishes on all rungs, whereas $n_+$ is large on every other rung. This leads to the peak at $k=\pi/2$ and no other strong features, as shown in the third panel of Fig.~\ref{fig:sf_1}. In the CI phase, particles sit on every other rung, so that $n_+$ is finite only for those rungs; $n_-$ is also finite only on these rungs, but in addition oscillates in sign. Therefore, we find a peak in $S_+$ at $k=\pi$ and $S_-$ at $k=\pi/2$, $k=3\pi/2$. As discussed above, the I phase is most easily understood by introducing an overall charge degree of freedom in the CI phase, but keeping the alternating structure. Consequently, the peak in $S_+$ is strongly reduced compared to those in $S_-$, cf. the second panel of Fig.~\ref{fig:sf_1}. The bottom panel, corresponding to the supersymmetric point, illustrates the proximity of this point to the ordered phases.

In Figure~\ref{fig:dens}, the correlation functions $\langle n_{0,\up} n_{i,\up} \rangle$ and $\langle n_{0,\up} n_{i,\dw} \rangle$
for antiperiodic boundary conditions are shown, where the diameter of the circles indicates the strength of the respective correlation. This corresponds to the density structure that would be obtained if translational symmetry was broken. The upper panel shows the system in the CI phase. Clearly, the density structure is consistent with the expectations and reminiscent of what is shown in the lower panel of Fig.~\ref{fig:gapped_phases}. The center panel shows the density correlation in the I phase. Clearly, density correlations are weaker in this phase, which is expected due to the gapless mode. The Ising order is nevertheless still discernible. Finally, the lower panel shows the system in the C phase. For sites sufficiently far away from the 0'th rung, the symmetry between the upper and lower chain is restored, which is expected for the Ising disordered phase. Charge ordering is also clearly visible, with every other rung showing much higher occupation.

\subsubsection{Doping away from quarter filling} \label{sct:resB3}

Finally, we study the behavior of the system when doped away from quarter filling. When the system is doped away from quarter filling by inserting holes, these holes become itinerant in both of the fully gapped phases, giving rise to a gapless charge mode. The central charge in these phases therefore becomes $c=1$. The line of KT transition separating the two Ising ordered phases disappears in this case and only two phases remain. These are separated by a line of Ising transitions, which appear as a line of $c=3/2$ transitions within a Luttinger liquid phase.

\begin{figure}
  \includegraphics{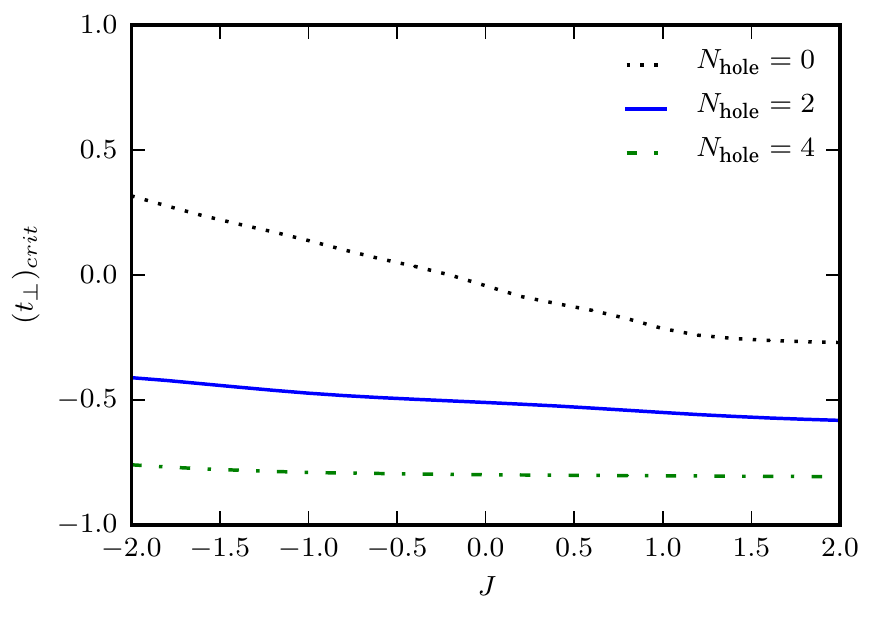}
  \caption{Location of the Ising transition in the $t_\perp$-$J$ phase diagram for doping away from ground state filling. The transition is determined from the peak in the central charge as $t_\perp$ is varied for fixed $J$. Calculations are performed for $L=24$ with the number of particles $N=L/2-N_{hole}$. \label{fig:ising_pd} }
\end{figure}

 In Fig.~\ref{fig:ising_pd}, we show a numerically obtained phase diagram for $L=24$ with 0, 2 and 4 holes. The phase diagram was established by measuring the central charge for a fixed system sizes $L=24$ on a fine parameter grid. The phase boundary can then be observed as a line with $c=1.5$, whereas we have $c=1$ in the rest of the phase diagram (for the number of holes $N \neq 0$) as all phases have at least a gapless charge mode. While the location of the phase boundaries is not of much interest here, this confirms that the Ising and charge disordered phase can be realized in a wide parameter range by doping the system away from quarter filling.

\subsection{Supersymmetry-preserving perturbation} \label{sct:zstag}

\begin{figure}
  \includegraphics{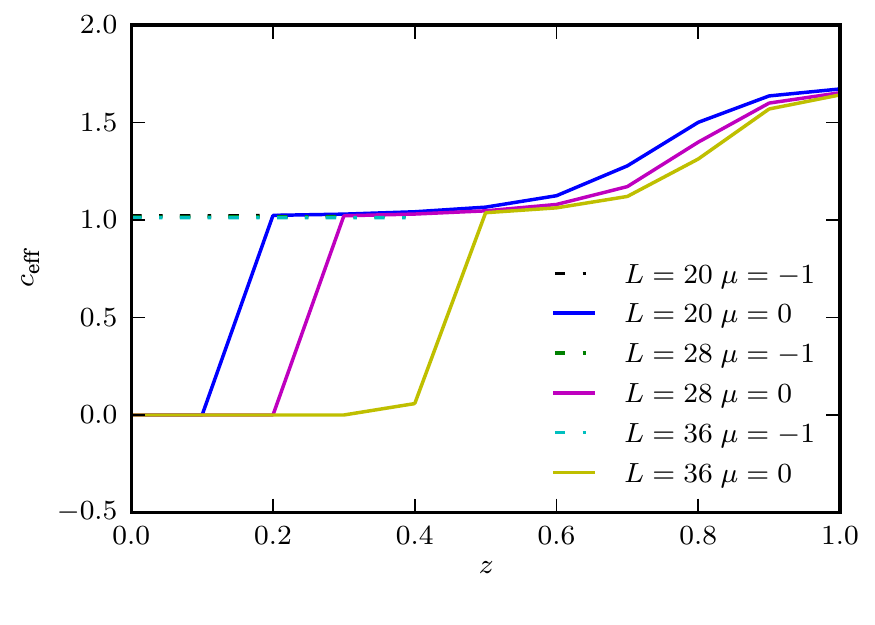}
  \caption{Effective central charge for systems perturbed away from the isotropic supersymmetric point by staggering on two sublattices, cf. Eqn.~\eqnref{eqn:z}. The chemical potential $\mu$ is only added on sites on sublattice $S_2$. \label{fig:cc_z} }
\end{figure}

The second supersymmetric minimal model contains a supersymmetry-preserving relevant operator that induces a flow towards the first supersymmetric minimal model with $c=1$.~\cite{leaf1993} We can therefore obtain further evidence for the conjectured continuum theory by identifying a microscopic perturbation that corresponds to this operator in the field theory. A possible starting point to find such a perturbation is the fact that the supersymmetric model on the \emph{chain} is described by the first supersymmetric minimal model with central charge $c=1$ in the continuum limit. We can therefore look for a perturbation that tunes the system between the square ladder and the chain, i.e. we look for a perturbation that meets the following criteria: i) The perturbation should preserve supersymmetry on the lattice. ii) For a large enough strength of the perturbation, the ground state of the perturbed ladder system coincides with that of the chain, and in particular is critical with $c=1$. iii) The Ising sector becomes gapped when turning on the perturbation, but the charge sector remains gapless; the unperturbed limit and the chain limit are connected by a line of critical systems with $c=1$.

Supersymmetry-preserving perturbations can be constructed using the staggering discussed briefly in Section~\ref{sct:model}. 
The specific staggering divides the lattice into two sublattices $S_1$ and $S_2$ with
\begin{equation}
S_2 = \lbrace (4n-3,\up), (4n-1,\dw) | n=1,\ldots,L/4 \rbrace,
\end{equation}
where $(i,\up)$ ($(i,\dw)$) indicates the upper (lower) site of the $i$'th rung. The supercharge is then defined as
\begin{equation} \label{eqn:z}
Q^\dagger = \sum_{i \in S_1} c_i^\dagger P_i + z \sum_{i \in S_2} c_i^\dagger P_i.
\end{equation}
Note that $S_2$ coincides with the occupied sites in the bottom panel of Fig.~\ref{fig:gapped_phases}.
For $z=1$, the system coincides with the homogeneous ladder system studied before. In the limit $z = 0$, there is no hopping term between the two sublattices and the particle number on each sublattice becomes a good quantum number. One can easily determine that one possible ground state of the system with periodic boundary conditions has all particles on sublattice $S_2$, which is a non-entangled product state. Since the total three-fold ground state degeneracy must remain unchanged under staggering, there must be two ground states where all particles are on sublattice $S_1$. This sector is equivalent to a chain with length $L'=3L/2$, where $L$ is the length of the ladder, and filling 1/3. The two-fold ground state degeneracy is then consistent with the degeneracy of the chain model, and we know this state to be critical with $c=1$ and described by the first $\mathcal{N}=2$ supersymmetric minimal model. We can easily select the ground state sector corresponding to a critical chain by adding a chemical potential on sublattice $S_2$ such that particles on that sublattice are penalized. This form of staggering therefore meets the first two criteria mentioned above, namely that it preserves supersymmetry and that in a limiting case ($z=0$), the model coincides with the exactly solvable model on the chain.

For $z > 0$, the number of particles on each sublattice is not a good quantum number and we expect the ground states to mix. We therefore have to resort to numerical simulations to determine whether the system remains gapless with $c=1$ for all $z \in [0,1]$.
We numerically calculate the effective central charge for values of $z$ between the limiting cases $z=0$ and $z=1$. Our results are shown in Fig.~\ref{fig:cc_z}. The solid lines show results without chemical potential. Clearly, the central charge drops from $c=3/2$ to $c=1$ away from the $z=1$ point. For small values of $z$, it becomes increasingly difficult to obtain the critical ground state because the DMRG calculation is biased towards the low-entanglement solution of particles sitting only on the $S_2$ lattice. While this is a true ground state only at $z=0$, it is very competitive in energy also for $z > 0$. This is remedied by adding a negative chemical potential on $S_2$ (i.e., energetically penalizing particles that sit on $S_2$) for small values of $z$ (dashed lines). This gives $c=1$ also for small values of $z$. Our data are therefore consistent with a $c=1$ phase for $0 \leq z < 1$, which is connected to the supersymmetric model on the chain.

\section{Conclusion}

We have studied in great detail a supersymmetric lattice model at and in the vicinity of a multicritical point. Despite being amenable to many analytic approaches, the continuum theory had not been firmly established prior to our work. This is mostly due to strong finite-size effects, which are caused by an unusual marginal operator present in the superconformal field theory. The marginal operator breaks Lorentz invariance, but preserves part of the supersymmetry. We demonstrate that the finite-size corrections are strongly suppressed in the entanglement entropy. This allows us to use entanglement entropy as a powerful probe for the system's properties. In particular, we show with careful finite-size extrapolation that the central charge of the model is $c=1.5$, which is the value expected for the second superconformal minimal model.

Augmenting entanglement entropy results with calculations of correlation functions, we are able to establish a three-parameter phase diagram realizing four different phases adjacent to the supersymmetric point. The supersymmetric point appears as multicritical point in this phase diagram. We can relate the adjacent phases to the Ising and charge sector present in the candidate superconformal field theory, providing additional strongly supportive evidence. Finally, we study a particular supersymmetry-preserving perturbation which establishes a relation of our model to the well-understood case of the chain, which is described by the first superconformal minimal model.

Our results underline that the entanglement entropy is a powerful probe for many-body systems since it can easily be obtained using the density-matrix renormalization group and since it shows very robust features.

The c=3/2 superconformal field theory, which we identify as the continuum theory, is also found as the continuum theory of related supersymmetric models.\cite{fendley2003-1,hagendorf2012} One of these models\cite{hagendorf2012} exhibits both the $\mathcal{N}=1$ and the $\mathcal{N}=2$ supersymmetry of the continuum theory on the lattice. Therefore coupling to the marginal operator that we have found is excluded in this model. It would be interesting to see if relating this model to our model provides a way of identifying the marginal operator on the lattice. Ideally, such an identification would allow one to tune away the finite-size effects due to the marginal operator.\cite{affleck1989} More generally, it might be possible to obtain qualitative verifications of the RG equations that we derived by studying supersymmetry preserving perturbations of the lattice model.

Finally, we note that it has been suggested that multicriticality is a generic feature of the supersymmetric lattice models and that the extensive ground state entropy in two dimensional systems is indicative of this feature.\cite{huijse2008,galanakis2012} The results of our paper may be seen as a small step in the very ambitious program of identifying multicriticality in these two dimensional models. An obvious next step is to carry out a similar analysis and explore the phase diagram by perturbing away from the supersymmetric point for a one dimensional model with an extensive ground state entropy such as the zig-zag ladder.\cite{huijse2008,HuijseThesis}


\section*{Acknowledgements}

It is a pleasure to thank Jan de Boer, Pasquale Calabrese, Massimo
Campostrini, Paul Fendley, Robert Konik, Andreas Ludwig, John
McGreevy, Bernard Nienhuis, Subir Sachdev, Ari Turner, Thomas Quella and Guifre Vidal and Herman Verlinde for
helpful discussions. We also thank the Aspen Center for Physics
and the NSF under Grant No. 1066293 for hospitality during the
early stages of this work and the Kavli Institute for Theoretical
Physics and the NSF under Grant No. NSF PHY11-25915 for
hospitality at the final stages of this work. Part of the
calculations were performed on the Brutus cluster at ETH Zurich.
The DMRG code was developed with support from the Swiss platform
for High-Performance and High-Productivity Computing (HP2C) and
based on the ALPS libraries.~\cite{bauer2011-alps} L.H. acknowledges
funding from the Netherlands Organisation for Scientific Research
(NWO). E. B. was supported by the National Science Foundation
under Grants DMR-0757145 and DMR-0705472.

\appendix
\section{Exactly soluble limits} \label{app:sollimits}

Here we review the results of the unpublished work of Ref.~\onlinecite{vdNoort} on the two soluble limits of the lattice model for hardcore spinless fermions on a 2-leg ladder. The starting point is the supersymmetric model on the square ladder given by the Hamiltonian in Eqn. (\ref{eqn:Ham}). To this Hamiltonian the perturbations $t_\perp$ and $J$ given in Eqn. (\ref{eqn:pert}) are added. The soluble limits are $J\to \infty$, $t_\perp \geq-1$ and $t_\perp=-1$, $J\geq -2$, respectively. In the following we assume that we tune the chemical potential such that the ground state is always at quarter filling.

Let us first consider the limit $J\to \infty$, where we recover an Ising model. Large $J$ implies a large energy penalty for particles to be on adjacent rungs. At quarter filling we can thus consider the low-energy subspace spanned by the configurations with a particle on every other rung. This subspace falls apart into two disconnected sectors with particles on all even/odd rungs. Projecting out the empty rungs, we are left with a Ising degree of freedom at each site representing a particle on the upper (lower) leg as $\uparrow$ ($\downarrow$). The hopping along the rung directly translates into a spin flip. The 3-body term is always trivial in this low energy subspace, whereas the 2-body repulsion for two particles to be two sites apart on the same leg simply translates into a ferromagnetic repulsion between two neighboring Ising spins. Finally, there is another contribution to this ferromagnetic interaction from second order perturbation theory, where a particle virtually hops to a neighboring rung and back. Putting all this together the effective Hamiltonian for large $J$ reads
\begin{eqnarray}
H_{J\to \infty} = \sum_{i=1}^{L/2} \Big[ (1+t_\perp) \sigma^x_i +(\frac{1}{2}+\frac{1}{2+J})\sigma^z_i \sigma^z_{i+1}\Big], \nonumber
\end{eqnarray}
where $\sigma^x$ and $\sigma^z$ are the usual Pauli matrices. It follows that for large $J$ there is an Ising transition at $t_\perp=-\frac{1}{2}+\frac{1}{2+J}\approx -1/2$.

We now turn to the limit $t_\perp=-1$. In this limit there is no rung hopping, and since two particles cannot hop past each other the Hilbert space falls apart into many disconnected sectors.~\cite{cheong2009-1} From the large $J$ limit, however, we know that for $t_\perp$ smaller than the critical value, the system is in an Ising ordered phase. In the Ising ordered phase the ground state is dominated by the configuration in which a particle on the upper leg is followed by a particle on the lower leg and vice versa. It follows that for $t_\perp=-1$ the ground state must be in the sector where particles on the upper and lower leg alternate. In this sector the attractive 3-body term and the repulsive 2-body term for particles on the same leg add up to zero for all configurations. The only remaining interaction term is the 2-body repulsion diagonally across a plaquette between a particle on the upper and a particle on the lower leg. It follows that for the effective Hamiltonian, we do not have to distinguish between particles on the upper and lower leg. Using a Jordan-Wigner transformation we can map the configurations of occupied and empty rungs to spin configurations. The nearest-neighbor hopping along the legs then translates into a nearest-neighbor spin exchange term and the 2-body repulsion into a spin-spin interaction. The effective Hamiltonian for $t_\perp=-1$ reads
\begin{eqnarray}
H_{t_\perp=-1} = 4 \sum_{i=1}^{L} \Big[ S^+_i S^-_{i+1} + S^-_i S^+_{i+1}+\frac{2+J}{4} S^z_i S^z_{i+1}\Big], \nonumber
\end{eqnarray}
where $S^z=\sigma^z/2$ and $S^\pm=(\sigma^x\pm \imath \sigma^y)/4$. The effective Hamiltonian is the well-known XXZ Heisenberg model. The continuum theory is the free boson. The KT transition between the charge ordered and charge disordered phase coincides with the $SU(2)$ symmetric point. Since this symmetry is also present in the lattice model the critical point is easily identified to be at $J=0$. In principle one can consider perturbing away form the $t_\perp=-1$ limit, but we will not do so here.~\cite{vdNoort}

\section{Relating the field theory to microscopics}\label{app:microscopic}

In this section, we present a microscopic way to understand the origin
of the effective field theory (Eq. \ref{eqn:L}). This provides a
way to relate the physical observables to the continuum fields $\Phi$,
$\psi$.

The microscopic Hamiltonian, Eq. \ref{eqn:Ham}, describes a one-dimensional
system of particles with strong repulsive interactions. Due to the
infinite nearest-neighbor repulsion, the statistics of the particles
is unimportant when considering properties such as the spectrum and the density-density correlations. This is since exchanging particles is impossible. In what follows,
we will imagine that the particles are \emph{bosons}, and derive an
effective low-energy theory for two interacting bosonic chains. This
procedure involves relaxing the hard-core constraints and replacing
them by ``soft core'' repulsive interactions, and then extrapolating
to the limit of strong interactions. Our derivation is based on the
assumption that in the low-energy limit, the physics of the soft-core
model matches that of the original model. This is supported by comparing
the phase diagram of the effective model (see below) to the numerical
results of Sec. \ref{sct:results} and the exactly solvable limits (Appendix \ref{app:sollimits}).

We begin from an effective bosonized Hamiltonian describing two chains
of interacting bosons, given by

\begin{equation}
H=H_{0}+H_{\mathrm{int}}.\label{eq:H_2chains}
\end{equation}
The free part of the Hamitonian describes two identical Luttinger liquids:

\begin{equation}
H_{0}=\sum_{\alpha=1,2}\frac{v}{2\pi}\int dxK\left[\left(\partial_{x}\phi_{\alpha}\right)^{2}+\frac{1}{K}\left(\partial_{x}\theta_{\alpha}\right)^{2}\right],
\end{equation}
where $v$ and $K$ are the sound velocity and Luttinger parameter,
respectively, and $\phi_{\alpha}$, $\theta_{\alpha}$ ($\alpha=1,2$) are canonical
fields satisfying $\left[\phi_{\alpha}(x),\partial_{x}\theta_{\alpha'}(x')\right]=i\pi\delta_{\alpha\alpha'}\delta(x-x')$.
The boson annihilation operator on chain $\alpha$ is given by $\psi_{\alpha}\sim\sqrt{\rho_{\alpha}}e^{i\phi_{\alpha}}$,
where $\rho_{\alpha}$ is the density operator of chain $\alpha$.

The interacting part of the Hamiltonian, $H_{\mathrm{int}}$, is most conveniently
written in terms of center-of-mass and relative variables, $\theta_{\pm}=\theta_{1}\pm\theta_{2}$,
$\phi_{\pm}=\left(\phi_{1}\pm\phi_{2}\right)/2$. Notice that these
fields satisfy the same commutation relations as the original fields
$\phi_{\alpha}$, $\theta_{\alpha}$. In terms of $\{\phi_{\pm}$, $\theta_{\pm}\}$,
$H_{\mathrm{int}}$ is given by

\begin{align}
H_{\mathrm{int}} & =\int dx\left(V\cos(2\theta_{-})-t_{\perp}\cos(2\phi_{-})\right)\nonumber \\
 & +\int dxV_{u}\cos(4\theta_{+})+\dots.\label{eq:Hint}
\end{align}
$V$, $t_{\perp}$, and $V_{u}$ represent inter-chain density-density
interactions, inter-chain hopping, and inter-chain Umklapp scattering,
respectively. We neglect terms involving higher harmonics, which are
less relevant than those in Eq. (\ref{eq:Hint}).

When $V=t_{\perp}$, the odd sector can be mapped into a transverse
field Ising model at criticality.\cite{Shelton1996} The Ising order/disorder
fields are identified as $\sigma\sim\sin\theta_{-}$, $\mu\sim\cos\phi_{-}$.
If $V\ne t_{\perp}$, the odd sector becomes gapped. $V>t_{\perp}$
($V<t_{\perp}$) corresponds to the ordered (disordered) phase of
the Ising model, respectively.

We can now see the relation between the bosonic two-chain model (\ref{eq:H_2chains})
and the effective field theory of Eq. (Eq. \ref{eqn:L}). The bosonic field
$\Phi$ is related to the center of mass degree of freedom: $\Phi=2\sqrt{2}\theta_{+}$.
The fermionic (Ising) sector in Eq. (\ref{eqn:L}) is equivalent to the odd sector
of our two-chain model. As discussed in Sec. \ref{sct:pd}, the supersymmetric
model corresponds to tuning the even and odd sectors to their critical
points simultaneously. Supersymmetry breaking perturbations open a mass gap in either the charge sector, the Ising sector, or both.

We can now relate the microscopic density to the operators of the
field theory. The density operator in chain $\alpha$ is given by

\begin{equation}
\rho_{\alpha}=\rho_{0}+\frac{1}{\pi}\partial_{x}\theta_{\alpha}+\frac{1}{a}\cos\left(2\pi\rho_{0}x+2\theta_{\alpha}\right)+\dots\label{eq:rho}
\end{equation}
Here, $a$ is the lattice constant, and the average density is $\rho_{0}=1/(4a)$,
corresponding to a quarter filled system. The $\dots$ represent higher
harmonics. It is convenient to define even and odd combinations of
the densities, $\rho_{\pm}=\rho_{1}\pm\rho_{2}$. In terms of the
even and odd fields, we get the following expression for $\rho_{\pm}$:

\begin{align}
\rho_{+} & =2\rho_{0}+\frac{1}{\pi}\partial_{x}\theta_{+}+\frac{2}{a}\cos\left(\theta_{-}\right)\cos\left(2\pi\rho_{0}x+\theta_{+}\right)\nonumber \\
 & +\frac{2}{a}\cos\left(2\theta_{-}\right)\cos\left(4\pi\rho_{0}x+2\theta_{+}\right)+\dots\label{eq:rho_plus}\\
\rho_{-} & =\frac{1}{\pi}\partial_{x}\theta_{-}+\frac{2}{a}\sin\left(\theta_{-}\right)\sin\left(2\pi\rho_{0}x+\theta_{+}\right)+\dots\label{eq:rho_minus}
\end{align}
Notice that in the odd sector, either $\cos\left(2\theta_{-}\right)$
or $-\cos\left(2\phi_{-}\right)$ are relevant (corresponding to the
ordered and disordered phases of the Ising model). So either $\theta_{-}$
or $\phi_{-}$ are pinned to the minima of the respective cosine terms.
In either case, $\langle\cos\left(\theta_{-}\right)\rangle=0$. Therefore,
the first harmonic in $\rho_{+}$, proportional to $\cos\left(\theta_{-}\right)$,
does not contribute to the long-range correlations, and can be dropped.
On the other hand, $\langle\cos\left(2\theta_{-}\right)\rangle\ne0$
in either the disordered or the ordered phase (since this term appears
in the Hamiltonian). Hence, the long-range correlations of $\rho_{+}$
contain a smooth part and an oscillatory part with wavevector $2Q\equiv4\pi\rho_{0}=\pi/a$.
The long-range correlations of $\rho_{-}$ are dominated by an oscillatory
part with period $Q=\pi/2a$, and the amplitude is proportional to
the Ising order parameter $\sigma\sim\sin\theta_{-}$. The expected
periodicities of the density modulations match the ones that appear
in the numerically obtained density structure factors (see Figs. \ref{fig:sf_1}, \ref{fig:dens}).
We conclude that Eq. (\ref{eq:rho_plus},\ref{eq:rho_minus}) give
the long-range components of the density operators in terms of the
continuum fields $\Phi\sim\theta_{+}$, $\sigma$.

\section{Supersymmetry preserving marginal operator}\label{app:susy}

We show that the multicritical theory of the free boson and the free fermion coupled to a marginal operator preserves part of the supersymmetry present at the fixed point. At the multicritical point this theory has an  $\mathcal{N}=(2,2)$ supersymmetry generated by the supercharges $G_L^{\pm}\equiv\psi_L \exp(\pm \imath \sqrt{2} \Phi_L)$ and $G_R^{\pm}\equiv\psi_R \exp(\mp \imath \sqrt{2} \Phi_R)$. At the fixed point these supercharges clearly satisfy $\partial G_R^{\pm}=0$ and $\bar{\partial} G_L^{\pm}=0$. Indeed the supercharges are the conserved currents associated to supersymmetry.

To show that part of the supersymmetry is preserved when we perturb the fixed point theory by coupling to a specific marginal operator, we show that a linear combination of the supercharges, $G_L^{\pm}$ and $G_R^{\pm}$, is still conserved at least perturbatively in the coupling constant. That is, we find new supercharges $\tilde{G}^{\pm}=(\tilde{G}_L^{\pm},\tilde{G}_R^{\pm})$ that satisfy 
\begin{equation}
\bar{\partial} \tilde{G}_L^{\pm}+\partial \tilde{G}_R^{\pm} =0 \nonumber
\end{equation} 
to first order in the coupling. As we will see, the new supercharges are no longer purely holomorphic or anti-holomorphic. This is because the marginal operator breaks Lorentz invariance and therefore couples the left- and right-moving sectors. It follows that the perturbed theory only preserves $\mathcal{N}=2$ supersymmetry.

The procedure to identify conserved currents away from the fixed point is nicely explained in Ref. \cite{Cardy88}. Consider a free theory with a conserved current, $J(z)$. Upon coupling the free theory to an operator $O(z,\bar{z})$ with coupling constant $g_O$, one find to first order in $g_O$ that
\begin{equation}\label{perturbativederivative}
\bar{\partial} J(z) = g_O A(z,\bar{z})+\dots 
\end{equation}
where $A(z,\bar{z})$ is the term in the OPE of $J$ with $O$ with coefficient $(z-z_1)^{-1}$:
\begin{equation}
J(z) O(z_1,\bar{z}_1) = \dots + \frac{1}{z-z_1} A(z_1,\bar{z}_1)+\dots \nonumber
\end{equation}
For a derivation of this result we refer to Ref. \cite{Cardy88}. The conclusion is that there is a conserved current if $A$ can be written as total derivatives with respect to $z$.

For the case under consideration, we perturb the fixed point theory by adding the following interaction terms to the Lagrangian
\begin{eqnarray}
\mathcal{L_{\rm{int}}}&=& -\imath \lambda O_{\lambda} (z,\bar{z}) + g O_{g} (z,\bar{z}), \nonumber\\
O_{\lambda} (z,\bar{z}) &=&  (\partial_x \Phi)  \psi_R \psi_L =(\imath \partial \Phi_L-\imath \bar{\partial} \Phi_R) \psi_R \psi_L \nonumber\\
O_{g} (z,\bar{z}) &=& \cos[\sqrt{2} \Phi] = \frac{1}{2}( e^{\imath \sqrt{2} \Phi}+ e^{-\imath \sqrt{2} \Phi}), \nonumber
\end{eqnarray}
where $\Phi=\Phi_L+\Phi_R$. For the OPEs of the supercharges with the marginal operators we find
\begin{widetext}
\begin{eqnarray}
 G_L^{\pm} (z) O_{\lambda} (w,\bar{w})  &\sim& \dots -\frac{\imath}{4 \pi}  \frac{1}{z-w}\psi_R \left(  \partial \Phi_L-\bar{\partial} \Phi_R \right) e^{\pm \imath \sqrt{2} \Phi_L } +\dots  \nonumber\\
G_R^{\pm} (\bar{z}) O_{\lambda} (w,\bar{w}) &=& \dots  + \frac{\imath}{4 \pi} \frac{1}{\bar{z}-\bar{w}}  \psi_L \left(  \partial \Phi_L - \bar{\partial} \Phi_R  \right) e^{\mp \imath \sqrt{2} \Phi_R } +\dots \nonumber\\
G_L^{\pm} (z) O_{g} (w,\bar{w}) &=& \dots\pm \imath \frac{1}{\sqrt{2}}  \frac{1}{z-w} \psi_L  \partial \Phi_L e^{\mp \imath \sqrt{2} \Phi_R }  +\dots \nonumber\\
G_R^{\pm} (\bar{z}) O_{g} (w,\bar{w}) &=&\dots \mp \imath \frac{1}{\sqrt{2}} \frac{1}{\bar{z}-\bar{w}} \psi_R \bar{\partial} \Phi_R e^{\pm \imath \sqrt{2} \Phi_L } +\dots   \nonumber
\end{eqnarray}
where all the operators one the r.h.s. are at $(z,\bar{z})$. We can ignore the more singular terms and regular terms, because they vanish upon angular integration~\cite{Cardy88}. Using these equations and the result (\ref{perturbativederivative}), we find that to first order in $\lambda$ and $g$ we have
\begin{eqnarray}
\bar{\partial} G_L^{\pm} &=& -\frac{\lambda}{4\pi} \psi_R ( \partial \Phi_L - \bar{\partial} \Phi_R) e^{\pm \imath \sqrt{2} \Phi_L} +  \frac{g}{2} (\pm \imath \sqrt{2} )\psi_L \partial \Phi_L  e^{\mp \imath \sqrt{2} \Phi_R}  \nonumber\\
 &=& \partial \left(  \frac{\lambda}{4\pi} \frac{\pm \imath}{ \sqrt{2}} \psi_R e^{\pm \imath \sqrt{2} \Phi_L} \right)  + \frac{\lambda}{4\pi}  \psi_R \bar{\partial} \Phi_R e^{\pm \imath \sqrt{2} \Phi_L} +  \frac{g}{2} (\pm \imath \sqrt{2}) \psi_L \partial \Phi_L e^{\mp \imath \sqrt{2} \Phi_R} \nonumber\\
\partial G_R^{\pm} &=&  \frac{\lambda}{4\pi} \psi_L (\partial \Phi_L - \bar{\partial} \Phi_R) e^{\mp \imath \sqrt{2} \Phi_R}  +  \frac{g}{2}  (\mp \imath \sqrt{2} )\psi_R  \bar{\partial} \Phi_R e^{\pm \imath \sqrt{2} \Phi_L} \nonumber\\
 &=&   \bar{\partial} \left( \frac{\lambda}{4\pi}  \frac{\mp \imath}{\sqrt{2}} \psi_L e^{\mp \imath \sqrt{2} \Phi_R}  \right)  + \frac{\lambda}{4\pi} \psi_L \partial \Phi_L e^{\mp \imath \sqrt{2} \Phi_R}  +  \frac{g}{2}  (\mp \imath \sqrt{2}) \psi_R  \bar{\partial} \Phi_R e^{\pm \imath \sqrt{2} \Phi_L}  \nonumber
\end{eqnarray}
\end{widetext}
We note that $\bar{\partial} G_L^{\pm}$ contains a total derivative with respect to $z$ and then two terms that cannot be written as total derivates. However, the latter two terms also appear in $\partial G_R^{\pm}$. It follows that a linear combination of the left- and right-moving supercharges leads to a conserved current if these terms cancel. For the terms to cancel, it suffices that the determinant of the matrix
\begin{eqnarray}
\left( \begin{array}{cc}  \frac{\lambda}{4\pi}  & \frac{g}{2}  (\mp \imath \sqrt{2}) \\ \frac{g}{2} (\pm \imath \sqrt{2}) & \frac{\lambda}{4\pi} \end{array}\right) \nonumber
\end{eqnarray}
vanishes. This leads to 
\begin{eqnarray}
\left( \frac{\lambda}{4 \pi}\right)^2- \left( \frac{g}{\sqrt{2}}\right)^2=0 \Rightarrow \lambda= \pm 2 \sqrt{2}\pi g .\nonumber
\end{eqnarray}
We thus find that for $\lambda= \pm 2 \sqrt{2}\pi g$ there are two conserved currents associated to an $\mathcal{N}=2$ supersymmetry because a linear combination of $\partial G_L^{\pm}$ and $\bar{\partial} G_R^{\pm}$ is equal to total derivatives. For $\lambda=2 \sqrt{2}\pi g$ the current conservation relation reads
\begin{eqnarray}
\bar{\partial} \tilde{G}_L^{\pm}+\partial \tilde{G}_R^{\pm}=0, \nonumber
\end{eqnarray}
with
\begin{eqnarray}
\tilde{G}_L^{\pm} &=&G_L^{\pm} + \frac{g}{2} \psi_L e^{\mp \imath \sqrt{2} \Phi_R}  \nonumber\\
\tilde{G}_R^{\pm}&=& \mp \imath G_R^{\pm} \mp \imath \frac{g}{2} \psi_R e^{\pm \imath \sqrt{2} \Phi_L} . \nonumber
\end{eqnarray}
Finally, we note that the supercharges generating the $\mathcal{N}=(1,1)$ supersymmetry are not conserved for any non-zero value of $\lambda$.

\bibliographystyle{apsrev4-1}
\bibliography{refs}

\end{document}